\documentclass[%
 reprint,
 amsmath,amssymb,
 aps,
]{revtex4-2}

\usepackage{cleveref}

\usepackage{graphicx}% Include figure files
\usepackage{dcolumn}% Align table columns on decimal point\usepackage{bm}% bold math
\usepackage{xcolor}
\begin{document}

\preprint{ }

\title{Using holographic microscopy to measure the effect of confinement on crowding agents in lipid vesicles}% Force line breaks with \\

\author{Yaam Deckel}%
\thanks{These authors contributed equally}
\affiliation{%
 School of Chemistry, Australian Centre for Astrobiology, and ARC CoE for Synthetic Biology, UNSW Sydney, NSW 2052, Australia}

\author{Lauren A. Lowe}%
\thanks{These authors contributed equally}
\affiliation{%
 School of Chemistry, Australian Centre for Astrobiology, and ARC CoE for Synthetic Biology, UNSW Sydney, NSW 2052, Australia}

\author{Siddharth Rawat}%
\thanks{These authors contributed equally}
\affiliation{%
 School of Chemistry, Australian Centre for Astrobiology, and ARC CoE for Synthetic Biology, UNSW Sydney, NSW 2052, Australia}

\author{Matthew Turner}%
\affiliation{School of Physics, The University of Sydney, NSW 2006, Australia}

\author{James Luong}%
\affiliation{School of Chemistry, The University of Sydney, NSW 2006, Australia}

\author{Anna Wang}
\email{anna.wang@unsw.edu.au}
\affiliation{%
 School of Chemistry, Australian Centre for Astrobiology, and ARC CoE for Synthetic Biology, UNSW Sydney, NSW 2052, Australia}
 

\begin{abstract}
    The hydrodynamic effects of macromolecular crowding inside cells is often studied \textit{in vitro} by using polymers as crowding reagents. Confinement of polymers inside cell-sized droplets has been shown to affect the diffusion of small molecules. Here we develop a method, based on digital holographic microscopy, to measure the diffusion of polystyrene microspheres that are confined within lipid vesicles containing a high concentration of solute. We apply the method to three solutes of varying complexity: sucrose, dextran, and PEG, prepared at $\sim$7 \% (w/w). We find that diffusion inside and outside the vesicles is the same when the solute is sucrose or dextran that is prepared below the critical overlap concentration. For polyethylene glycol, which is present at a concentration higher than the critical overlap concentration, the diffusion of microspheres inside vesicles is slower, hinting at the potential effects of confinement on crowding agents.
\end{abstract}

\maketitle

%%%%%%%		 Keywords			%%%%%%%    

%	 Please provide up to 5 keywords. These will be displayed at the end of the manuscript, before the reference section.

\section*{Introduction}
\label{introduction}

Macromolecular crowding is an unavoidable feature of living systems: The interior of cells is extremely crowded, averaging 200 mg/mL protein by mass -- twice more concentrated than egg white~\cite{ellis_join_2003}. This crowding has ramifications for solutes, and both membraneless and membrane-bound compartments. Macromolecular crowding can shift biomolecular condensate equilibria, altering the thermodynamics of condensate formation~\cite{minton_influence_2001}. Crowded cell interiors affect membrane remodelling and the steady-state shape of the membrane~\cite{fujiwara_generation_2014}. Crowding can also impact both active and passive transport within cells, including reaction rates~\cite{mourao_connecting_2014, zhou_macromolecular_2008, wilcox_overlap_2020}.

The hydrodynamic effect of macromolecular crowding inside cells is often studied by using polymers as crowding agents. Most studies of macromolecular crowding focus on the impact on nanoscale diffusion. To study this, techniques such as fluorescence correlation spectroscopy~\cite{watanabe_cell-size_2018} and fluorescence recovery after photobleaching~\cite{jawerth_protein_2020, harusawa_membrane_2021} are used to monitor the diffusion of a population of fluorophores. 

Optical microscopy can be used to understand the impact of crowding on larger objects, by using microspheres as probes. One challenge with using micrometre-sized colloidal particles as tracers is that they can diffuse out of plane during image acquisition, unless the samples are extremely viscous~\cite{jawerth_protein_2020}. When using conventional microscopy methods, the particle can be challenging to localise when it diffuses out of focus, leading to captured particle trajectories being truncated. For example, a 1-µm-diameter particle in water will on average diffuse 3 µm in 10 seconds. Imaging bulk concentration profiles~\cite{collins_nonuniform_2019} enables diffusion to be analysed, however, the trajectories for individual particles are difficult to capture. 

\begin{figure}
\begin{center}
\includegraphics[width=8.6cm]{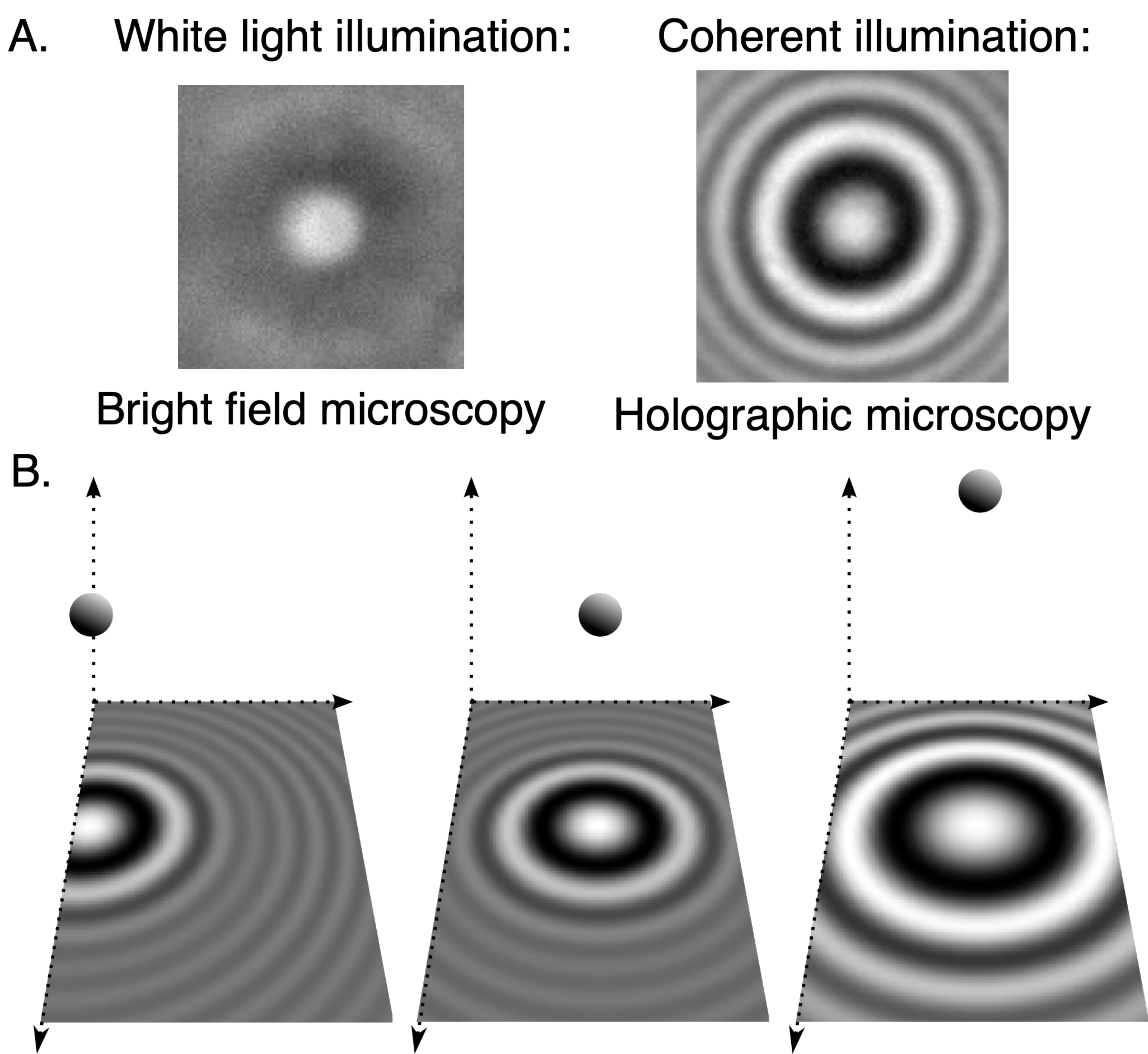}
\caption{(A) Colloidal particles that are out of focus under white light illumination (bright field microscopy) have a higher-contrast diffraction pattern when illuminated with coherent light. This imaging mode is also known as holographic imaging. Images of 1-$\mu$m-diameter polystyrene particles are shown. (B) When simulated 1-$\mu$m-diameter polystyrene particles translate in-plane, their hologram centres also translate. When particles translate out-of-plane, the interference fringes of the hologram change, still enabling the particle to be tracked.}
\label{fig1}
\end{center}
\end{figure}

One way to capture longer particle trajectories is to use a coherent light source in place of the bright field light source (Fig.~\ref{fig1}A), hence forming an in-line holographic microscope. The coherence length of the light enables the scattering from out-of-plane objects to be recorded as an interference pattern (i.e. hologram). As particles move in three dimensions the centroid and shape of the hologram change (Fig.~\ref{fig1}B), enabling the trajectory to be extracted by comparison to a generative model~\cite{martin_-line_2022}.

This work builds on the recent results by Watanabe and Yanagisawa, who found that the confinement of polyethylene glycol (PEG) inside cell-sized droplets affects the diffusion of tracer molecules, with diffusion inside the confined droplets being faster than diffusion in bulk~\cite{watanabe_cell-size_2018}. Their results showed that PEG prepared above the critical overlap concentration likely forms heterogeneous structures when subjected to confinement and ageing, enabling small tracers to diffuse faster within the voids.

Here we use digital holographic microscopy to measure the diffusion of polystyrene (PS) microspheres confined within lipid vesicles containing a high concentration of solute. Digital holographic microscopy has previously been used to measure the 3D diffusion of colloidal tracers in bulk solution~\cite{cheong_holographic_2009, wang_using_2014}, both to measure the microrheological properties of the system, and also to use the diffusion data to accurately size particles. The tracking precision using holography can be as good as 1 nm in each of the three dimensions.~\cite{martin_-line_2022} Here, we develop a methodology to encapsulate single polystyrene tracers within giant unilamellar vesicles (GUVs) that are suitable for holographic imaging, then measure the diffusion of the tracers both inside and outside the GUVs. Our results suggest that diffusion of micrometre-scale tracers may be slowed down by confinement if the polymer is restructured by the confinement.

\section*{Results and Discussion}
\label{results_discussion}

\subsection*{Methodology and design considerations}

To study this phenomenon, we needed to devise a protocol to generate GUVs that encapsulated PS beads as microrheological tracers. We chose to use a self-assembly method~\cite{kindt_bulk_2020, lowe_methods_2022} for several reasons. First, there is no oil phase involved, and thus no oil droplets wetting the PS particles or membrane to bias the measurement results. Whilst oleic acid is an oil, we deprotonate the oil with NaOH to form an aqueous micelle phase to prepare the GUVs.~\cite{kindt_bulk_2020} Importantly, the viscosity of an oil phase external to a lipid layer could impact the hydrodynamics because of non-zero transmission of shear across lipid bilayers~\cite{amador_hydrodynamic_2021}. Second, the method enables the encapsulation of colloidal particles~\cite{kindt_bulk_2020}. Third, this self-assembly method results in GUVs that have the same solution inside and outside the vesicles~\cite{kindt_bulk_2020}, enabling straightforward comparison between confined and unconfined particles. Fourth, the lack of refractive index contrast between the vesicle's interior and exterior simplifies the image analysis by avoiding any artefacts from lensing.

In brief, oleate micelles (pH $>$ 10) are added to a buffered solution (Na-bicine pH 8.06) that results in the rapid rearrangement of the oleate/oleic acid into vesicle membranes. During this self-assembly process, some particles become encapsulated within the vesicles. From previous work, we know that a 5 mM solution of oleate/oleic acid vesicles can self-assemble into an extremely high yield of GUVs, whereas using higher than 20 mM of lipid generates a jammed solution of giant vesicles, some of which are vesicle-in-vesicle structures that are undesirable for this work because the presence of internal vesicles can complicate the interpretation of results~\cite{kindt_bulk_2020}. We therefore worked within this concentration range of 5-20 mM.

\begin{figure}
\begin{center}
\includegraphics[width=8.6cm]{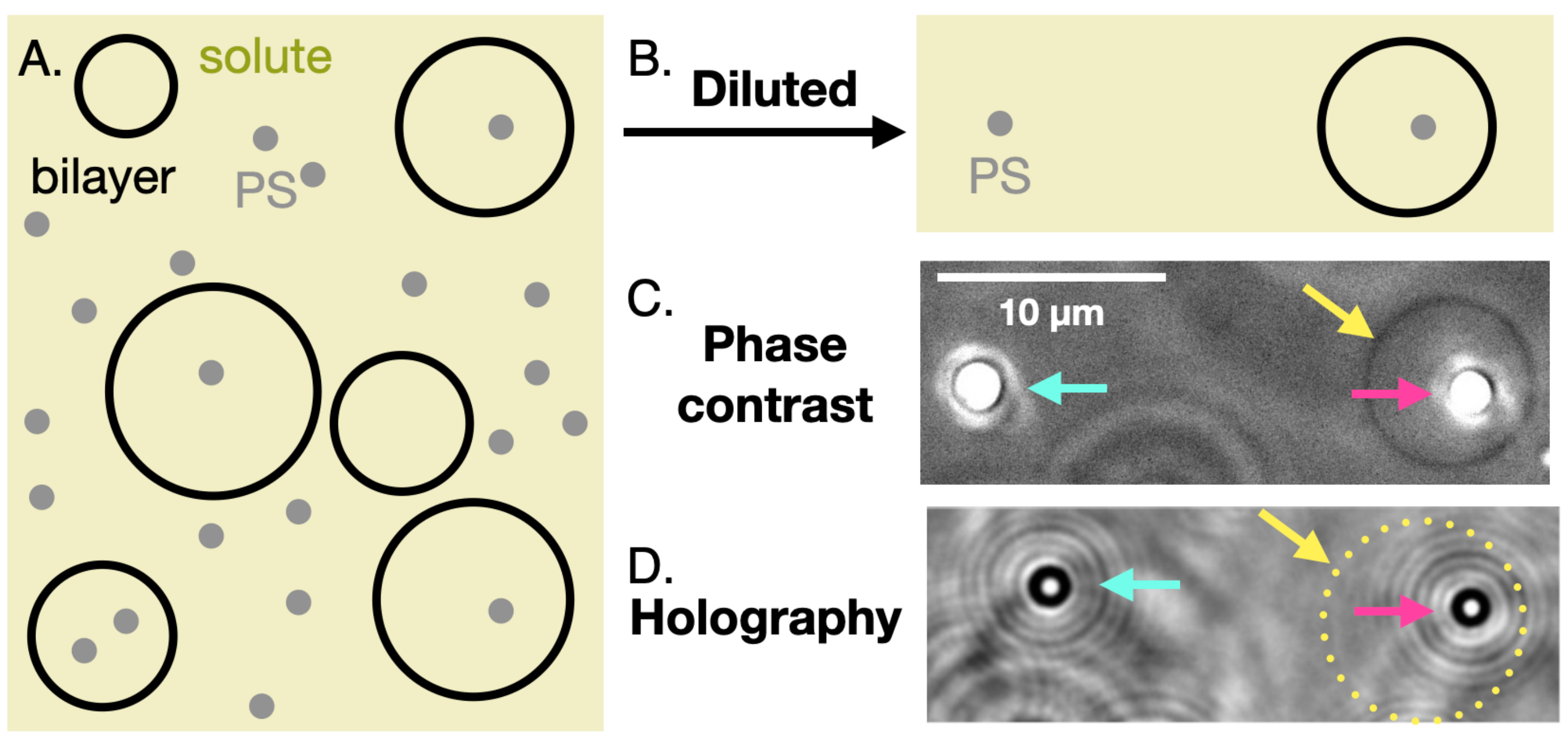}
\caption{(A) GUVs self-assemble in the presence of solute and polystyrene beads (PS). The concentrations of particles and lipid are optimised for GUVs containing just one particle. Most vesicles will have no particles inside, and some will have more than one. (B) If the sample is too crowded with particles, it is diluted. (C--D) The vesicle membrane is visible under phase contrast imaging (yellow arrow) but only barely so under holography (yellow arrow, dotted yellow line). Particles that are free (cyan arrow) and trapped (magenta arrow) can be identified using phase contrast microscopy, then tracked with holography.}
\label{fig2}
\end{center}
\end{figure}

While the self-assembly technique has previously been shown to encapsulate many colloidal particles at once~\cite{kindt_bulk_2020}, here we aimed to work in a dilute regime such that most vesicles would encapsulate either zero or one PS bead. This ensures that the hydrodynamic coupling between nearby particles does not complicate the data analysis. Assuming that the particle encapsulation process is random, and a vesicle diameter of 10 µm and a particle diameter of 1 µm, the volume fraction of particles required is no more than 1/1000. Because out-of-plane particles are still visible under holographic illumination, the presence of unwanted scatterers (such as too many unencapsulated particles) needed to be minimised~\cite{martin_-line_2022}. Owing to the high critical aggregation concentration of oleic acid relative to phospholipids~\cite{wang_lipid_2019}, the sample was not diluted below 250 µM oleic acid to avoid the dissolution of GUVs. This process resulted in samples containing a dilute suspension of unencapsulated PS, alongside a dilute suspension of GUVs that mostly encapsulated either one or no PS particles (Fig.~\ref{fig2}A-B).

Determining whether the particles were trapped inside vesicles, or free, was a challenge in holographic imaging mode. This is because the GUVs, having membranes that are only a few nanometres thick~\cite{wang_core-shell_2019}, do not scatter much light. The scattering from the membranes becomes even more challenging to detect in the presence of PS beads. We therefore used a two-step approach to determine which PS beads to image: 1) use phase contrast mode to find GUVs that contained one PS bead (Fig.~\ref{fig2}C), 2) use the holographic imaging mode to capture holograms of the particles diffusing. The process was repeated for particles that were not trapped inside GUVs.

\subsection*{Comparing diffusion of trapped and free particles}
\begin{figure}
\begin{center}
\includegraphics[width=8.6cm]{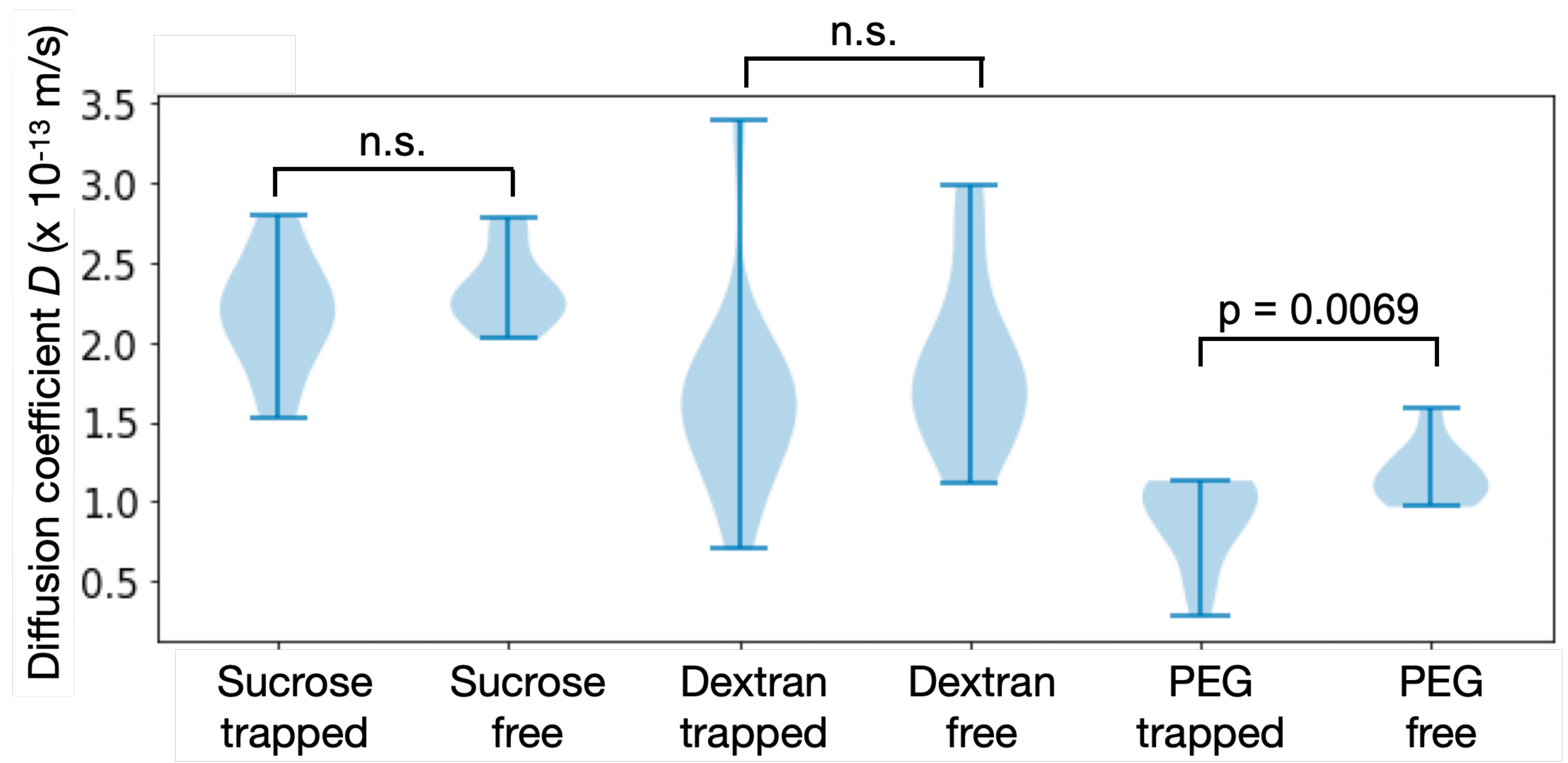}
\caption{Diffusion coefficients of trapped ($N$ = 10) and free ($N$ = 10) particles in 0.475 M sucrose solution are not significantly different.  Diffusion coefficients of trapped ($N$ = 16) and free ($N$ = 14) particles in 8.75 \% dextran solution are not significantly different. Diffusion coefficients of trapped ($N$ = 16) and free ($N$ = 10) particles in 9.5 \% PEG solution are significantly different. Violin plots show spread of data, with bars indicating full range of values measured.}
\label{fig3}
\end{center}
\end{figure}

As a negative control, we measured the diffusion of PS beads inside and outside lipid vesicles in a buffer containing 0.5 M sucrose. We found that diffusion is the same inside and outside the vesicles (Fig.~\ref{fig3}), even though the bead is only one order of magnitude smaller in diameter than the GUVs -- that is, the confinement of the PS bead inside the GUVs has a negligible effect on the PS bead's diffusion. This result concurs with that of Amador and coworkers, who found that fluid bilayers have almost complete flow transmission across the bilayer, with the drag force increasing only slightly near the bilayer~\cite{amador_hydrodynamic_2021}. We anticipate that the increase in drag would be even more minimal for oleic acid membranes, which are more fluid than phospholipid membranes~\cite{kindt_bulk_2020}, thereby imposing less drag on nearby particles. In other words, we would expect the proximity of the tracer to the membrane to have a relatively small impact on diffusion, because the external phase is identical to the internal phase and shear is transmitted across the bilayer.

We then measured the diffusion of PS beads inside and outside lipid vesicles in a buffer containing 8.75 wt\% dextran (average molecular weight 10 kDa). Wilcox and coworkers experimentally measured the critical overlap concentration ($c*$) of the same dextran to be approximately 25 wt\%~\cite{wilcox_overlap_2020}, and thus our preparation is considered to be in the dilute regime. Again, we found that diffusion is the same inside and outside the vesicles (Fig.~\ref{fig3}). This is in agreement with the work by Harusawa and coworkers, who saw that confinement within 20-µm-diameter droplets did not impact the diffusion of nanometre-sized tracers via interactions between the crowding agent and the membrane for low dextran concentrations~\cite{harusawa_membrane_2021}.

Finally, we measured the diffusion of PS beads inside and outside lipid vesicles in a buffer containing 9.5 wt\% PEG (average molecular weight 8 kDa) that was first `aged' for at least one week as per Watanabe and Yanagisawa~\cite{watanabe_cell-size_2018} then encapsulated using our method into GUVs. We follow Watanabe and Yanagisawa~\cite{watanabe_cell-size_2018} to calculate the critical overlap concentration for the PEG using \cref{eq1} for the radius of gyration $R_g$ (in nm) and \cref{eq2} for $c*$:

	\begin{equation} \label {eq1}
			R_g = 0.02 M^{0.58}
	\end{equation}

        \begin{equation} \label {eq2}
			c* = M N_A / (\frac{4 \pi} {3} R_g^3)
	\end{equation}
where $N_A$ is Avogadro's number, and $M$ is the molecular weight, giving $c*$ = 6.4 wt\% in our system. In other words, our PEG solution at a concentration $c$ = 9.5 wt\% is in the semidilute regime. The correlation length $\xi$ can be calculated by $\xi = R_g (c/c^*)^{-0.588} = 2.91$ nm,~\cite{kalwarczyk_motion_2015} which is much smaller than the size of the diffusive PS tracers.

We find that the diffusion for trapped particles is slower than for free particles (Fig.~\ref{fig3}). Watanabe and Yanagisawa hypothesised that the accelerated diffusion of their nanoscale tracers, at sub-millisecond time scales, was possible owing to the restructuring of the PEG at the sub-micrometre scale~\cite{watanabe_cell-size_2018}. Our results support that the confinement from lipid bilayers is only restructuring the crowding agent at these smaller lengthscales, because our trapped 1-µm-diameter tracers diffuse slower, and not faster, than in bulk. Indeed, it appears that despite the ageing of the PEG, confinement could be enhancing crowding for micrometre-scale objects, owing to potential interactions of the PEG with the membrane~\cite{harusawa_membrane_2021}. Another explanation could be that the restructuring of the PEG that leads to voids at the nanoscale results in a more viscous environment at the micrometre scale. Additional experiments with PEG at different concentrations, dextran at different concentration, and PEG and dextran and different molecular weights, can help conclusively demonstrate the effect of crowding on confinement.

These results show that holographic imaging can be used to perform microrheology in crowded environments, even when under confinement. It can be used as a complementary technique to fluorescence correlation spectroscopy~\cite{watanabe_cell-size_2018} or bulk concentration profile measurements~\cite{collins_nonuniform_2019}. The presence of fluorophores can result in unwanted interactions with the membrane~\cite{hughes_choose_2014}, or alterations to self-assembly processes~\cite{quinn_how_2015, wagele_how_2019} or diffusion processes~\cite{adrien_how_2022}. Thus, if fluorophore-membrane interactions or fluorophore-crowder interactions are prevalent, a scattering-based fluorophore-free method such as holography may be desired, particularly if the colloidal tracer can be passivated. 

In principle, this procedure can be extended to more complex encapsulated mixtures. Homogeneous mixtures of solutes are expected to behave the same optically as the PEG, dextran, and sucrose solutions in this manuscript. If the encapsulated material generates phase-separated droplets, lensing from the droplets may introduce artefacts when tracking the PS tracer. The chemical functionalisation of the tracer must also be optimised for localisation inside one phase or the other.

\begin{figure}
\begin{center}
\includegraphics[width=8.6cm]{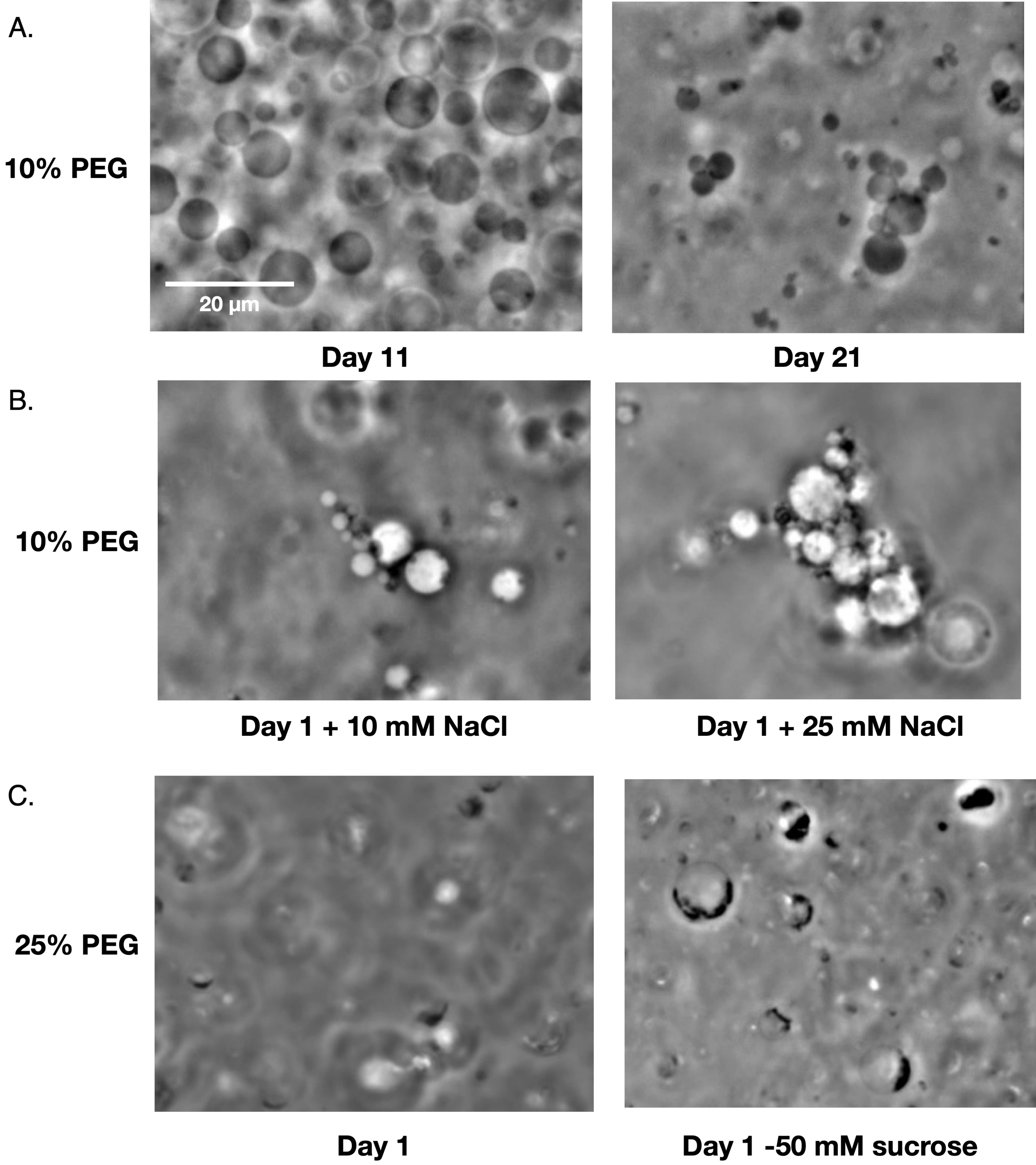}
\caption{Phase contrast images of oleic acid/oleate GUVs in buffer containing 50 mM sucrose, 100 mM bicine, diluted into buffers containing PEG and other solutes as indicated, to a final oleic acid/oleate concentration of 2 mM. (A) Vesicles in the presence of 10 \% PEG are still colloidally stable at Day 11, but begin aggregating by Day 21. They still retain their encapsulated solute (sucrose), as evidenced by (B) With slightly more salt, vesicles in the presence of 10 \% PEG are no longer colloidally stable even on Day 1. (C) Vesicles are immediately destroyed in the presence of 25 \% PEG. Even without sucrose, to minimise van der Waals interactions between vesicles~\cite{israelachvili_intermolecular_2011}, the vesicles start accruing debris from destroyed vesicles, and lose their encapsulated contents from Day 1.}
\label{fig4}
\end{center}
\end{figure}

One limitation is the concentration of solute that can be used in the presence of vesicles. High concentrations of macromolecules could result in significant depletion forces, which can cause undesired vesicle aggregation and/or depletion of the particles against vesicle membranes. We exposed GUVs to a range of solution conditions (Fig.~\ref{fig4}) and found that time, salt concentration, sucrose encapsulation, and PEG concentration can all affect vesicle colloidal stability, as expected from theory~\cite{israelachvili_intermolecular_2011}. We, therefore, recommend condition testing prior to preparing experiments for holographic characterisation. Alternatively, working in an extremely dilute-vesicle regime to minimise vesicle-vesicle interactions could prevent some unwanted aggregation. Alternate methods of preparing vesicles with an encapsulated particle, such as microinjection of particles into GUVs, or microfluidic methods where the external solution is exchanged to remove refractive index contrast between the inside and the outside of the vesicle, could also be used as an alternative to the GUV self-assembly method. The advantage of using these other techniques is that the encapsulation step is faster than the one day required for oleic acid GUV self-assembly, thereby enabling measurements of encapsulated polymer solutions that are yet to age.

A limitation of holography is that the tracer sizes required are large (at least 100 nm) relative to those used in fluorescence-based techniques (10-30 nm). However, these larger sizes are at the scale of cellular organelles, and could be used to understand the impact on such structures, especially if the crowded environment is heterogeneous. Examining a range of tracer sizes could be a fruitful direction for future work. Luby-Phelps and coworkers found that diffusion can be impacted in a size-dependent manner in cytoplasm extract~\cite{luby-phelps_hindered_1987}, revealing insights into the cytoplasm's structure. For larger particles or denser particles, particle buoyancy must be taken into account. In this work, the density between the tracer and the medium differed by no more than 0.03 g/cm$^3$, resulting in terminal sedimentation velocities of no more than 20 nm/s. Using the expression for mean squared displacement, $\Delta x^2(\tau) = 2D\tau$ where  $D$ is the diffusion coefficient for the most viscous system we studied ($\sim 10^{-13}$m/s), and $\tau$ is a lag time of 1 s, we find that the particle on average diffuses hundreds of nanometres per second. Diffusion thus dominates for these particular systems. Particles that are larger or denser could have significant sedimentation during the course of an experiment, in which case the particle would be probing the hydrodynamics of the near vicinity of the membrane instead of the internal aqueous lumen.

Holographic imaging has become simple to implement~\cite{giuliano_digital_2014, martin_-line_2022}, and hologram analysis tutorials are readily available online~\cite{barkley_holographic_2020, martin_-line_2022}. Future work using holographic imaging could probe how a series of tracers of different sizes are impacted by different crowders. While the tracers used in this study were 1-µm-diameter PS beads, smaller colloidal particles can be used if the material scatters more e.g. metal. With heterogeneity at microscopic length scales~\cite{watanabe_cell-size_2018} and concentration gradients in crowders~\cite{collins_nonuniform_2019} both leading to faster diffusion for smaller tracers, it may be interesting for future studies to investigate the impact of crowding on larger tracers using holography. Another subject of further study could be to examine a series of concentrations for a single crowding agent, to conclusively determine the concentrations for which confinement makes a difference to crowding.

\section*{Conclusion}
\label{conclusion}

Here we described the considerations required for trapping single colloidal particles inside GUVs that have a high concentration of solute both inside and outside of the vesicles. The resulting system was very low in contrast, with the vesicle only visible under phase contrast microscopy. The low scattering from the vesicles enabled the particles to be tracked using digital holographic microscopy. We then used these particles as microrheological tracers, and found that confinement within membranes may enhance the effects of crowding for micrometre-sized colloidal objects, if the crowding agent is in the semidilute regime.

\section*{Experimental}
\label{experimental}

\paragraph{Chemicals:}
Oleic acid ($\geq$99\%), bicine (99\%), poly(ethylene glycol) (average molecular weight 8000), 2.5 wt \% 1 µm diameter carboxylate-modified polystyrene latex particles (density 1.05 g/cm$^3$), dextran from Leuconostoc mesenteroides (average molecular weight 9-11 kDa), were purchased from Sigma-Aldrich and used as received. Sucrose and NaOH was purchased from Chem Supply. All water used was Millipore (18.2 M$\Omega \cdot$ cm).

\paragraph{Stock solutions:}
To make 0.1 M oleate stock, 5M NaOH (30 µL) and oleic acid (31.5 µL) were added to Milli-Q water (970 µL) before placing on the orbital shaker at 100 rpm (PSU-10i Grant Bio, UK) for 1 hour until clear. 

1 M sucrose stock solution, 10 \% w/v dextran solution, and 50 \% w/v PEG solution were also prepared by dissolving the solutes in Milli-Q water with overnight agitation of the tube on an orbital shaker. The PEG solution was aged for at least one week before being used for subsequent experiments.

1 M bicine stock solution was adjusted to pH 8.06 by the addition of NaOH.

\paragraph{Sucrose GUV solutions:}
Stock solutions were combined in a microcentrifuge tube to final concentrations of 9.5 mM oleic acid/oleate, 47.5 mM Na-bicine, 0.475 M sucrose, and 0.06 wt \% PS beads. The microcentrifuge tube was then agitated for 1 week on the orbital shaker at 100 rpm.

The resulting vesicles were diluted 1 in 20 into a dilution buffer containing 47.5 mM Na-bicine, 0.475 M sucrose. The density of the solution was approximately 1.06 g/cm$^3$.~\cite{heidcamp1996cell}

\paragraph{Dextran GUV solutions:}
Stock solutions were combined in a microcentrifuge tube to final concentrations of 5 mM oleic acid/oleate, 50 mM Na-bicine, 8.75 \% w/v dextran, and 0.06 wt \% PS beads. 

The resulting vesicles were diluted 1 in 20 into a dilution buffer containing 50 mM Na-bicine, 8.75 \% w/v dextran. The density of the solution was approximately 1.03 g/cm$^3$.~\cite{mach_density_1968}

\paragraph{PEG GUV solutions:}
Stock solutions were combined in a microcentrifuge tube to final concentrations of 9.7 mM oleic acid/oleate, 47.5 mM Na-bicine, 9.5 \% w/v PEG, and 0.003 wt \% PS beads. The microcentrifuge tube was then agitated for at least 1 week on the orbital shaker at 100 rpm. The density of the solution was approximately 1.02 g/cm$^3$.~\cite{gonzalez-tello_density_1994}

\paragraph{Imaging:}
5 $\mu$L of vesicle sample was sealed between a 22 x 22 mm coverslip and a 25 x 75 mm glass slide using silicone vacuum grease (Dow Corning). Vesicles were imaged by phase contrast and digital holographic microscopy using a 1.3 NA 100$\times$ objective (Nikon, Japan) on a TE-2000 inverted microscope (Nikon, Japan). Diascopic brightfield illumination was with a pT-100 LED, and holographic illumination was with a 660 nm mounted LED (Thorlabs) following the setup described by Giuliano and coworkers~\cite{giuliano_digital_2014}. Images were captured with a pco.edge 4.2 (PCO Imaging, Germany) at 80.5 Hz, with each trajectory lasting 1000 frames.

\paragraph{Hologram analysis:}
The holograms were analysed by iterative comparison to a Mie scattering model using the package HoloPy~\cite{barkley_holographic_2020}. As described by Martin and coworkers~\cite{martin_-line_2022}, this procedure can be used to localise the particle in each frame. In brief, Mie scattering is used to calculate the scattering from the PS tracer using information about its refractive index, size, and a guess for its three-dimensional location. This information is used to generate a modelled hologram. The modelled hologram is compared pixel-by-pixel to the data hologram, and the sum of the squared residuals is recorded as the cost function. By continually generating new holograms, a Levenberg-Marquardt algorithm then finds the best-fit values (refractive index, size, and three-dimensional location for the particle) that minimises the sum of the squared residuals.

\paragraph{Calculation of the diffusion coefficients:}
The best-fit particle coordinates were then analysed to extract the diffusion coefficient. The mean squared displacement MSD was calculated by $\Delta x^2(\tau) = <(x(t+\tau)-x(t))^2>$ for each lag time $\tau$ ranging from 12.4 ms to 124 ms. A straight line was then fit to the data of MSD against lag time. Because $\Delta x^2(\tau) = 2D\tau$, the slope of the fitted straight line is equal to $2D$. The diffusion coefficient $D$ can then be determined by dividing the value of the slope by 2. The correlated errors were not taken into account when calculating the average. For more detailed methodology that takes into account the correlated errors, we refer readers to the procedure by Wang, Dimiduk, and coworkers.~\cite{wang_using_2014}

\section*{Acknowledgements}
A.W. acknowledges support from the Australian Research Council (DE210100291), and the Human Frontier Science Program (RPG0029/2020 to A. W.).

\section*{Conflict of Interest}

The authors have no conflicts of interest to declare.

\bibliography{manuscript}

\end{document}